\documentclass[superscriptaddress, amsmath,amssymb, prl, twocolumn]{revtex4-1}
\usepackage{graphicx}
\usepackage{dcolumn}
\usepackage{bm}
\usepackage{hyperref}
\usepackage{color}
\usepackage{amsmath}
\usepackage{amssymb}

\begin{document}

\title{A  computational study  of a transversely propelling  polymer  and passive particles}
\author{K.~R.~Prathyusha}
\affiliation{School of Chemical and Biomolecular Engineering,
Georgia Institute of Technology, Atlanta, GA 30318}%
\email{krprathyusha@gmail.com}

\date{\today}

\begin{abstract}
Using Langevin dynamics simulations, we study a  system of transversely propelling filament and passive Brownian particles. We consider a polymer whose monomers experience a constant propulsion force perpendicular to the local tangent in the presence of passive particles undergoing thermal fluctuations in two dimensions. We demonstrate that the sideways propelling polymer can act as a sweeper to collect the passive Brownian particles, mimicking a shuttle-cargo system. The number of particles the polymer collects during its motion increases with time and finally saturates to a maximum number. Moreover, the velocity of the polymer decreases as the particles get trapped due to the extra drag they generate. Rather than going to zero, the velocity eventually reaches a terminal value close to the contribution from the thermal velocity when it collects the maximum load. We show that, apart from the length of the polymer, the propulsion strength and the number of passive particles are deciding factors for the maximum trapped particles.   In addition, we demonstrate that the collected particles arrange themselves in a triangular, closed, packed state, similar to what has been observed in experiments.

\end{abstract}

\pacs{Valid PACS appear here}

\maketitle

\section{I. Introduction}
  Active matter (\emph{motile}) comprises of individual units capable of converting energy to direct and autonomous propulsion and is capable of self-organization, pattern formation and coordinated motion ~\cite{marchetti-13}. In contrast to active particles, passive particles (\emph{non-motile}) only undergo Brownian motion due to random thermal fluctuations from the surrounding fluid and are exploited as reference systems to investigate the   formation of glasses or jammed state~\cite{glass-nature}. There is a growing interest in understanding the mixture of active and passive particles, as their collective effect is quite different from the dynamics of the system containing the corresponding single component alone \cite{cates-njp-active-passive-mixture-17,bechinger-SM-melting-of-colloidal-crystal-15,Ranni-SM-crystallizing-hard-sphere-14, AditiPrthiaphaseseparation-18,gompper-propogating-interfaceNJP-16,velocity-reversal-active-passive-Lowen-20,Cates-Phase-separation-aandp-mixturePRL-15,Ranni-glass-transition-towards-RCP-13}. 
   A key  application which benefits the understanding of  passive-active mixtures and the interaction between them  is  the  transportation of microscopic cargo~\cite{Ayusman-pnas-rod-shaped-motor-tracer-13,Wang-review-syntheticnanomachine-12,Ayusman-review-nanoscale-13}.  The controlled transport of nano/micro-scale cargo  is essential for many processes, such as directed drug delivery \cite{Ayusman-review-nanoscale-13}, micro-robotics \cite{Sitti-microrobotics-19,wang-micro-robots-formedicine-17}, bio-sensing \cite{sinwook-biosensing-20,wang-micro-robots-formedicine-17}, or removing unwanted particles from solutions \cite{pine-colloidal-docker-13,simmchen_micro_plasticremoval-19,janussweeper-21}.

Many cargo transport mechanisms are inspired by the  fundamental process inside the cell involving the transfer of molecules and molecular cargo over long distances mediated by motor proteins along actin filaments or microtubules \cite{science-motor-protein-review-07,Wang-review-syntheticnanomachine-12,vogel-review-molecular-shuttles-10}.  Apart from such motor proteins based molecular shuttles, recent attempts explored the possibility of natural micro swimmers, for instance motile bacteria or algae, as  cargo carriers~\cite{baceria-liposom-cargo-16,microoxenalgae-cargo-05,cargo-carryingbacteria-11,magnetobacteria-liposome-14,cargo-carryingSM-18,cargo-carrying-liquid-crystallineSM-15}. Although these motors are bio-compatible, their shortcomings, such as finite life span, time-sensitive motility, and its dependency on experimental conditions, can be overcome by synthetic micro/nano motors~\cite{baceria-liposom-cargo-16}. In artificial self-propelling systems, the agents  can be powered by various actuation mechanisms \cite{science-motor-protein-review-07}, and they appear in different shapes (spheres, rods, tubes) and sizes ~\cite{howse-Ramin-PRL-07, Ayusman-catalytic-motors-2006,schmidt-nanotubes2012,Ebbens-Ramin-PRE-12}.  The major challenge for the directed motion of the synthetic active particle is overcoming the Brownian motion, which randomizes  the particle orientations, and recent studies showed that  including external fields, or boundaries, can get over this difficulty \cite{Wang-review-syntheticnanomachine-12,bechinger-light-active-12,das-Ramin-boundaries-steering-NC-15}.  
 
  The advantage of the synthetic system is the ability to tune the interaction between the active coating of the motor and passive particles, which decide the specific behaviour of the mixture. An effective, attractive interaction between the passive and active particles is preferred for clustering and transport of cargo by the motor. Different kinds of interaction between motors and cargo have  emerged in the literature:  electro-static interaction  between the negatively charged  segment of the motor and the positively charged cargo beads
  ~\cite{positive-cargo-negative-rod-Ayusman-08,Ayusman-pnas-rod-shaped-motor-tracer-13},  bio-molecules with binding affinities coated/attached on cargo and motors~\cite{positive-cargo-negative-rod-Ayusman-08, simmchenDNA-motor-12,DNA-attached-Wang-11,tangenzymescience-robo-20}, magnetically driven motors~\cite{magnetic-interaction-wang-08}, cargo-motor assemblies with hydrophobic surface interaction~\cite{hydrophobicWang-13}.  Computational simulations or experiments  employed various  phoretic response to the gradient, to gain more understanding  of  complex dynamics during the cargo transport~\cite{schmidt2019light}. For the transport of cargo, when there is no surface functionalization,  alternate methods have been employed in the past, for instance, ferromagnetic microjet \cite{ferro-magnetic-jet-13}, and mobile microelectrodes \cite{label-freecargo-electricfield-18}. However, the studies based on the transport of non-specific cargo are still limited.  
   
     In the cargo transport process, for an efficient pick-up of cargo, it is preferable to have a self-propelling unit with a larger active surface area ~\cite{label-freecargo-electricfield-18, janussweeper-21}. When the propulsion direction is perpendicular to the short axis for a rod, a larger active "area" can be easily achieved. This is an effective way of collecting a large number of cargo compared to the conventional method ~\cite{Ayusman-pnas-rod-shaped-motor-tracer-13}. Even though tangentially propelled objects were the focus of the many research ~\cite{prathyusha_PRE-18, peruanibaer}, synthetic sideways-moving objects got much attention recently. The transverse propulsion of synthetic motor can be achieved either by coating two different metals on opposite sides of the fiber, creating an anisotropy~\cite{vrao-clasen-jphys-D-19,vutukuri-huck-SM16,christian-clasen-JMC-13},  surface generated gradient mechanisms \cite{Falko_fiberiod-2020} or external magnetic field~\cite{pietro-sagues-small10}. Recently we developed a  polymer model for such sideways-moving elongated objects, having connected beads with activity perpendicular to local tangents~\cite{prathyusha_softmatter-22}. A particularly promising application of the sideways propelling rod is its use in removing unwanted species from the solution~\cite{janussweeper-21}.   
     
     Inspired by the experiment~\cite{janussweeper-21}, we explore the possibility of employing a computational model of the transversely propelling polymer for passive cargo transport. Since the sweeper collects particles across an interface in experiments, one can assume a  two-dimensional motion of sideways moving polymer. 
 We use the polymer model for transversely propelling filaments where the active driving is perpendicular to the tangent~\cite{prathyusha_softmatter-22} of the polymer and moves on a substrate.   
  \begin{figure}
	\includegraphics[width=0.5\textwidth]{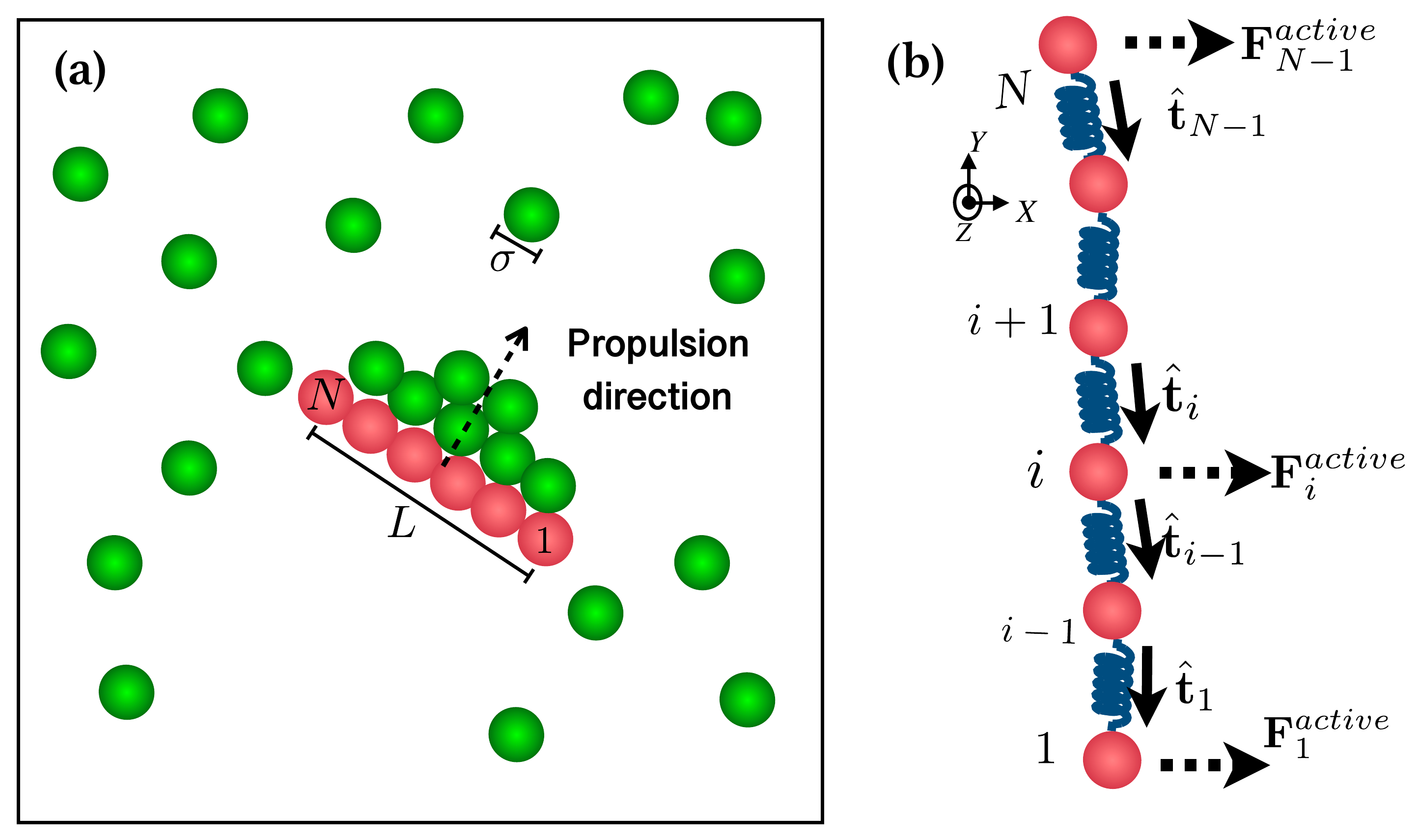}
	\caption{(a) Schematic of the system containing transverse polymer (green) and passive particles (green). (b) The model of transversely propelling polymer. The beads (monomers) are connected via harmonic springs, and bending potential between any connected triplets gives the stiffness to the polymer. Since the propulsion is perpendicular to the local tangent of the polymer,  the polymer adopts a motion along its short axis of the body. }
	\label{fig:polymer-schematic}
\end{figure}
For simplicity, we treat the non-specific cargos as passive Brownian particles undergoing thermal fluctuations. In this article, we study a mixture of a single transverse propelling polymer surrounded by randomly arranged Brownian particles using Langevin simulation. It is unclear from the experiments that how the aspect ratio/length of the polymer decides the number of particles collected by the polymer as it directly connects to the active "area" of collection.  In our simulations, we vary both the length of the propelling polymer and the number of passive particles. Our model ignores the effect of flow, but frictional coupling to the substrate is included. 
 
The paper is organized as follows. In Sec. II, we describe the coarse-grained model for  polymer having an active force perpendicular to it's local tangent,  model for Brownian particles and the simulation details. In Sec. III, we present and discuss the results of detailed Langevin dynamics simulations and describe how the sideways moving polymer can act as a sweeper. Finally, in Sec. IV, we summarize our main findings and  comment on future directions. 

\section{II.  Transversely propelling polymer model and simulation details}
We consider a $2D$ binary mixture of  $N_t$ passive Brownian spheres and a  polymer with $N$ self-propelling beads; see Fig.~\ref{fig:polymer-schematic}a.  The transversely propelling  polymer contains monomers  with co-ordinates ${\bf r}_i$ $(i =1, . . . , N)$, resembling an   "active sweeper polymer (rod/filament)".  Using the polymer model for the transverse filament, we can easily tune both the aspect ratio and stiffness of the polymer. The Brownian particles are modelled as passive beads  (with position co-ordinates ${\bf r}_{i}$ $(i =N+1, . . . , N+N_t$)), undergoing thermal fluctuations, mimicking the passive cargo particles.  
 The motion of the particles is governed by the following Langevin equation, 
 \begin{equation}
m_i\frac{d{\bf v}_i}{dt}={\bf F}^{WCA}_i +{\bf F}_i^{fric}+{\bf F}_i^{noise}+
 {\mathcal A}( {\bf F}_i^{stretch}+{\bf F}_i^{bend}+ {\bf F}_i^{active})  
 \label{eq:eqofmotion}
\end{equation}

The  first three terms (non-active force) in equation \ref{eq:eqofmotion} is experienced by both brownian beads and transverse polymer. The first term corresponds to the  steric interactions between  all the particles in the systems is modelled via Weeks-Chandler-Anderson potential \cite{weeks-chandler-71}, 
$U_{WCA}(r)=4\varepsilon\left[\left(\frac{\sigma}{r}\right)^{12}-\left(\frac{\sigma}{r}\right)^{6}+\frac{1}{4}\right],
\label{eqn:wca}
$ 
which vanishes  beyond any distance greater than $2^{1/6}\sigma$ and ${\bf F}^{WCA}_i= -\frac{\partial U_{WCA}(r)}{\partial r_j} {\bf \hat r}_{ij}$.
Here $\varepsilon$ measures the strength of the repulsive interaction, and $\sigma$ is the size of the bead, and both are set to $1$.  $r\equiv\left|\mathbf{r}_{ij}\right|=\left|\mathbf{r}_i-\mathbf{r}_j\right|$ is the distance between two beads. The interaction prevents the transverse polymer from crossing different segments of its chain as well as makes passive beads repulsive. Thus in our model, passive particles experience steric repulsion (see, Eqn.~\ref{eqn:wca})   between each other and with the active rod. We measure length in units of particle size $\sigma$, energies in units of $k_BT$, and time in units of $\tau=\sqrt{\frac{m\sigma^2}{k_BT}}$.

 ${\bf F}_i^{fric} = -\xi{\bf v}_i$ describes the drag force experienced by the particles when moving on the substrate with a  friction coefficient. ${\bf F}^{noise}_i$ is modelled as
white noise with zero mean and variance proportional to $ \sqrt{k_BTm \xi/\Delta t }$. 

Now, we discuss the last term in Eqn.~\ref{eq:eqofmotion}  containing contributions from three forces and is experienced by the polymer beads solely as $\mathcal A$ is set to 1 for the polymers. Since these interactions are absent for passive particles, we set ${\mathcal A}$  to zero for such particles. Each connected pair of the polymer interacts via a  harmonic potential \cite{kkremer-90}, $U_{stretch}(r)=k_b \big(r- R_0\big)^2 \label{eq:bond}$ (${\bf F}_i^{stretch}=-\nabla _{i} U_{stretch}$).  Here $R_0=1.0\sigma$ is the maximum bond length, and $k_b=4000~k_BT/\sigma^2$
is the bond stiffness. 
${\bf F}^{bend}_i$   is experienced by any connected triplet in the polymer, and it is derived from a potential, $U_{bend}= \kappa \big(\theta -\theta_0 \big)^2
 $ (${\bf F}^{bend}_i$= -$\nabla _{ijk} U_{bend}$). Here $\kappa$ is the bending stiffness and is related to the continuum bending stiffness, $\tilde \kappa$, as $\kappa \approx \tilde \kappa /2\sigma$,  $\theta$ is the angle between any consecutive bond vectors and $\theta_0$ is set to $\pi$. These parameters make the chain effectively inextensible with bond length $b\sim 1 \sigma$ and polymer length $ L =(N-1)b \sigma$.

The propulsion force ${\bf F}_i^{active}$ acts perpendicular to each local tangent of the polymer, resembling a  transversely propelled rod in the stiff limit~\cite{prathyusha_softmatter-22}. A  schematic of the polymer model is given in Fig.~\ref{fig:polymer-schematic}b, and all the beads of the polymer propel with the same force to model a 'sweeper' moving along the short axis of the body. The force experienced by the polymer beads is as follows, all  the beads of the backbone \{$i =2,3, N-1$\} of the polymer experience a force, having  contributions from two of its neighbours,
\begin{equation}
{\bf F}_i^{active}=  \hat{ \bf z} \times \frac{\hat{\bf t}_i +\hat{\bf t}_{i -1}}{2.0} {f_p} 
\label{activeforceb}
\end{equation}
 Where  $\hat {\bf t}_i= \frac{{\bf r}_i-{\bf r}_{i+1}}{|{\bf r}_i-{\bf r}_{i+1}|}$ is the 
local tangent between the beads, $i$ and $i+1$ and  $\hat{ \bf z}$ is the unit vector normal to the 2D plane of motion.  The active forces for the end beads ($i=1$ and $i=N$), as they  have only  one nearest neighbour,   are 
\begin{equation}
{\bf F}_{1}^{active}=  \hat{ \bf z} \times \hat{\bf t}_{1}  \alpha f_{p} \; \& \; \;
 \\
 {\bf F}_{N}^{active}=  \hat{ \bf z} \times \hat{\bf t}_{N-1} \alpha f_{p}
\label{activeforce-end}
\end{equation}
For the homogenous propulsion of all the beads, we keep the $\alpha = 1$. 
  
We vary the number of beads in the polymer as $N \in \{5, 10,15, 20\}$ and we chose the value of the $\kappa$ for each polymer length such that $\kappa/L k_BT \sim 1000$, making it a stiff polymer and it reduces the shape fluctuations of the polymer.  The temperature $T$ is set to $0.1$ so that effect of noise is less. The magnitude of active force vary in the range, ${f_p=\{0.01-100\}~ k_BT/\sigma}$ and  box size is $300\sigma \times 300 \sigma$.  Equation of motion \ref{eq:eqofmotion}, is integrated with time-step $\Delta t = 10^{-4} \tau $  using LAMMPS~\cite{splimpton-95}, incorporating the active force in the Eqns.~\ref{activeforceb} and \ref{activeforce-end}.  We place passive particles randomly in the simulations box along with the transverse polymer. For better averaging, we performed many simulations  started from  different  initial conditions. \cite{janussweeper-21} Since there is no effective, attractive interaction between the particles, the propulsion force, length of the sweeper and the number of passive particles are deciding factors for the maximum load the sweeper collects, and we present our results in the next session.

  \begin{figure}
	\includegraphics[width=0.5\textwidth]{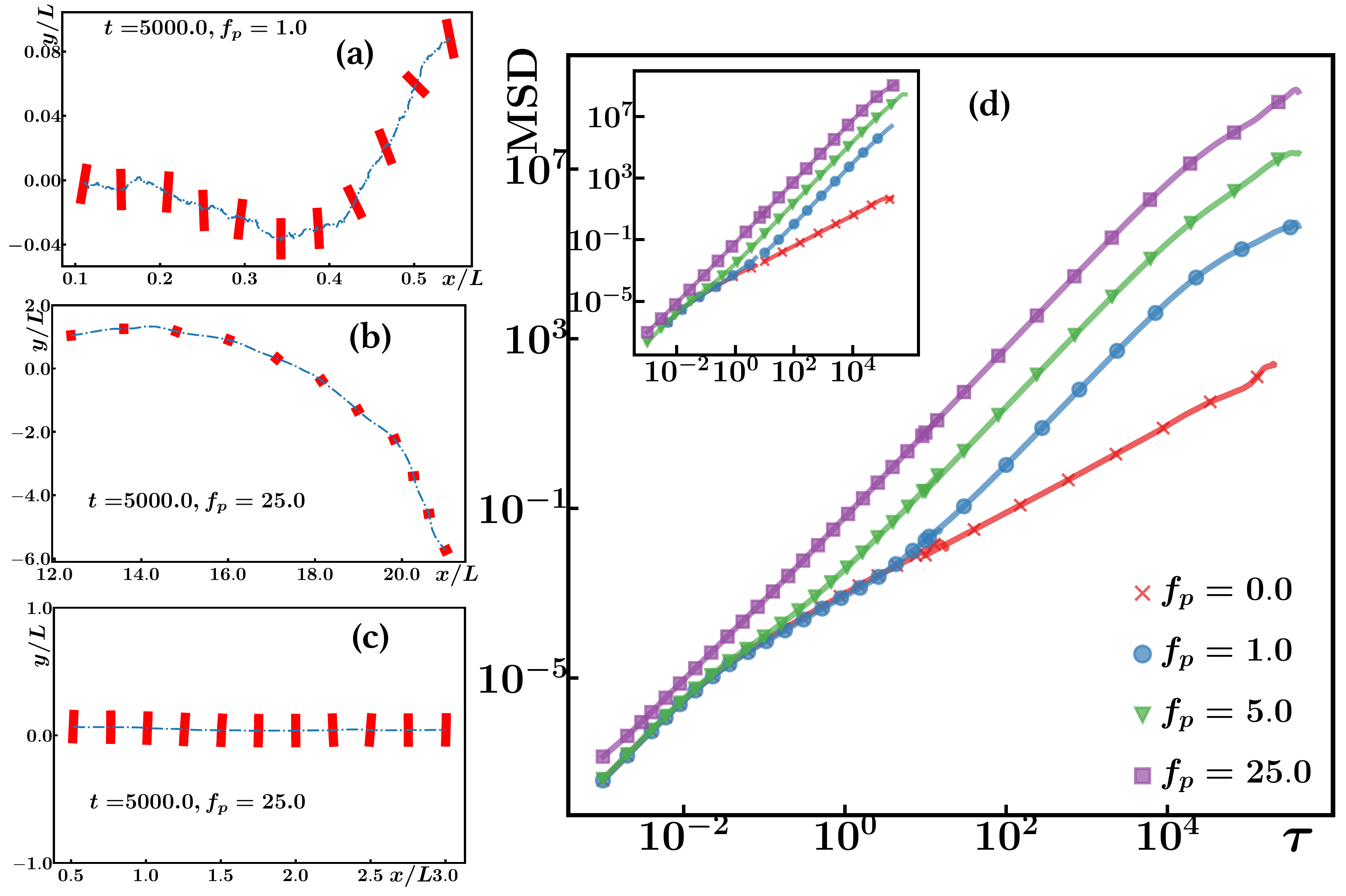}
	\caption{ (a) Trajectory of rods for different length and propulsion forces: (a) $N=5, f_p=1.0$, (b) $N=5, f_p=25.0$, (c)  $N=25, f_p=25.0$ (d) Mean squared displacement of the polymer ($N=5, \xi=0.1$) exhibit four different regime in the absence of tracer particles. Inset shows MSD for longer filament ($N=10$). 
	 }
	\label{fig:msd-no-load}
\end{figure}

 \section{III. Results}
  \section{Freely moving transversely propelling filament}
  \begin{figure}
	\includegraphics[width=0.4\textwidth]{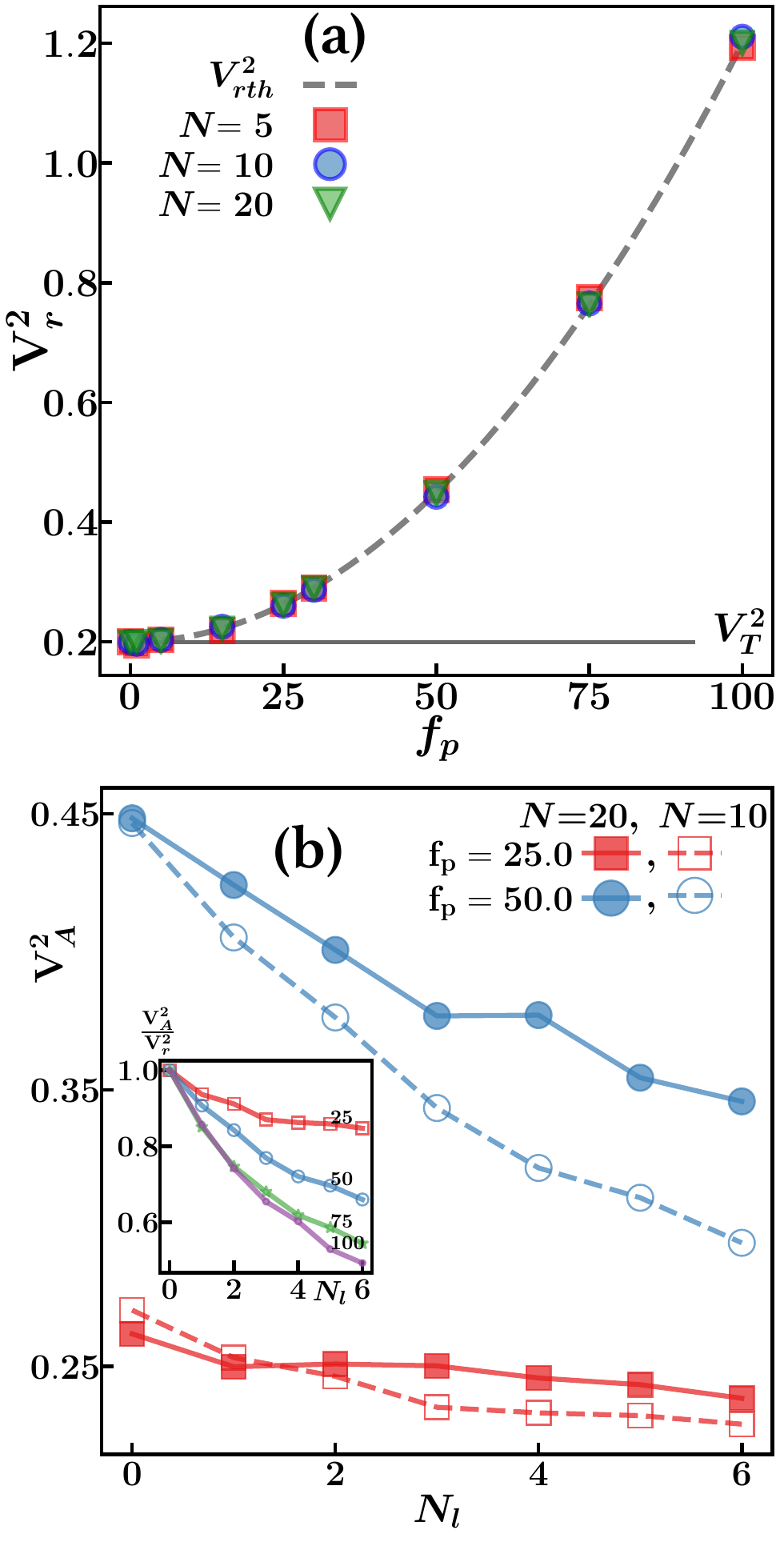}
	\caption{ (a) Mean squared velocity of the freely moving rod for different polymer length. The contribution from the thermal noise, $V_T^2$ is plotted for comparison. At higher values of $f_p$, contribution from activity dominates (b) Reduction of the mean squared velocity  when the rod pushes $N_l$ beads. Filled symbols are for $N=20$ and unfilled symbols  are for the case of $N=10$.  In the inset, we plot how the velocity of the rod ($N=10$) reduced w.r.t to the velocity of the freely propelling rod. The values of $f_p$  are given. Even though absolute value of $V_r^2$ is higher for larger $f_p$, the relative reduction is higher for larger values of $f_p$ when it pushes the same number of particles compared to a slow moving rod.  }
	\label{fig:vcm-no-load}
\end{figure}

To begin with, we examine the dynamics of the rod in the absence of any passive beads ($N_t=0$), and we consider stiff ($\kappa/k_BT >> L$)  transversely propelling filament. Since the strength of the noise is minimal, we expect all the monomers of the polymer to move identically, leading to a translational motion along the short body axis.  Fig.~\ref{fig:msd-no-load}(a), (b), and (c) describe the trajectory of the short and long filaments for different propulsion strengths. At a low value of the $f_p$, the short polymer exhibits random motion, see (Fig.~\ref{fig:msd-no-load}(a)). When the propulsion force is much higher than the thermal noise, the directed movement of short and long polymer for a long time is observed, see Fig.~\ref{fig:msd-no-load}(b)\&(c). This suggests that the motion behaviour of the polymer is highly dependent on the aspect ratio and propulsion force. It indicates that high propulsive force or a long polymer is preferred for a directed motion for a long duration. To quantify the motion we can determine ${\textrm {MSD}}=\langle|{\bf r}_{cm}(t+\tau)-{\bf r}_{cm}(t)|^2\rangle$ and is presented in the Fig.~\ref{fig:msd-no-load}(d). Studying the mean square displacement reveals different power-law regimes depending on the propulsion and length of the polymer. The major regimes are  (i) initial ballistic, (ii)~intermediate diffusive and (iii)~super diffusive and (iv)~diffusive regime. The initial ballistic regime appears in all the propulsion forces and polymer length, and it is due to the inertial effects. When the strength propulsion is much smaller than the effect of noise, the polymer motion shows a transition from ballistic to diffusive behaviour at ($m/\xi$) ($f_p=0,0.5$). At slightly longer time scales and at intermediate propulsion force, the diffusive regime shift to an active ballistic regime at $D/2V_r^2$  due to the directed motion of the polymer. Here, $D= k_BT/\xi $ is the diffusion coefficient, $V_r$ is  the c.o.m. velocity of the rod, $V_r=|\frac{1}{N}\sum_i^N(v_{ix}^2+v_{iy}^2)|$ and  $v_{ix}$ and $v_{iy}$ are  $x$  $\&$ $y $ velocity component of the $i^{th}$ monomer. Finally, a long-time diffusive regime with an enhanced diffusion coefficient appear due to the de-correlation of the initial orientation of the polymer. The time scale associated with this transition,  for a stiff polymer scales as $\xi L^3/k_BT$. Thus, the long-time diffusive regime is difficult to capture in our simulation for the long polymer; see insect of  Fig.~\ref{fig:msd-no-load}(d) for $N=10$. Similar plots are available for a different value of drag and propulsion force set in reference~\cite{prathyusha_softmatter-22}.
 
 Unlike in overdamped simulation, where the velocity of the particle is not an observable, underdamped simulations consider the time evolution of the velocity along with the particle trajectory. Since in Eqn.\ref{eq:eqofmotion}, the pairwise interactions (conservative) do not contribute to the center of mass velocity of the rod,  the mean squared velocity of the center of mass of the rod has only two contributions, $V_{rth}^2 = V_T^2+V_{P}^2$. Where  $V_T^2$ is the mean-squared thermal velocity in the absence of the propulsion force, and the equipartition theorem relates the velocity fluctuations to the solvent temperature in $2$ dimension as, $V_T^2=2k_BT/m $. $V_P^2$  is the contribution arising from the propulsion force, and it takes the form, $V_p^2=(f_p/\xi)^2$,  if polymer conformation is straight. Thus $V_{rth}^2$ takes the following form, 
 \begin{equation}
 V_{rth}^2 = 2k_BT/m+(f_p/ \xi)^2.
 \end{equation}

 In Fig.~\ref{fig:vcm-no-load}(a), we plotted the measured $V_r^2$ as a function of propulsion strength $f_p$  for different polymer lengths and $V_{rth}^2$ is plotted for comparison (the dashed line). It is clear from the figure that measured value of the mean squared velocity, $V_r^2$  and theoretical value, $V_{rth}^2 $ are in good agreement.  Since the all the polymer beads propel with same constance force, velocity of the rod does not get affected by the polymer length. For the higher values of  $f_p$, the propulsive forces dominate the contribution from the thermal noise. 
  
\section{Mixture of a transversely propelling filament and passive particles}

Next, we consider the case of a transversely propelling polymer in the presence of $N_t$  passive particles. It has to be noted that in our minimalistic model, there is no explicit attraction between the passive particles and the self-propelling rod. As there is no attractive interaction like in experiments (due to a self-induced electric field arising from the electrophoretic mechanism),  the attachment of passive particles to the rod depends on their mutual proximity.
 This means the sideways rod collects the nearby particles randomly during its motion. Initially, we examine the case $N_t< N $, and in the situation, the number of passive beads collected and pushed by the rod, $N_l=N_t$.  Here, $N_l$ particles fall within the cut-off $r_c = 2^{(1/6)} \sigma$ from any of the individual beads of the rod or any bead collected by the rod.   The propulsion of the polymer overcomes the thermal fluctuations and drag force exerted by the passive particles, and it carries them during the motion. Thus,  even without any specific attractive interactions between them, the high propulsion of the rod can cause passive particles to get stuck near the rod and get transported when it moves. The polymer velocity decreases with increasing number of captured spheres due to an increase in the resistance of the microrod-cargo aggregate in agreement with the experiments \cite{janussweeper-21, solovev2010magnetic-velocity}. In Fig.~\ref{fig:vcm-no-load}(b), we plot the mean squared velocity of the rod (different lengths and $f_p$) as a function of the number of loaded particles, $N_l$. The reduction in velocity for short polymer ($N=10$, unfilled symbols) and long polymer ($N=20$, filled symbols) follow similar trends for a given $f_p$. In the inset, we plot how the velocity of the rod is reduced when it pushes the particles, to the velocity of the freely propelling rod ($N_l=0$). The velocity of the slow-moving rods decreases to a constant value as it might have reached the maximum load it can afford to push, even with very few particles (see also the next section).

 \begin{figure}
	\includegraphics[width=0.45\textwidth]{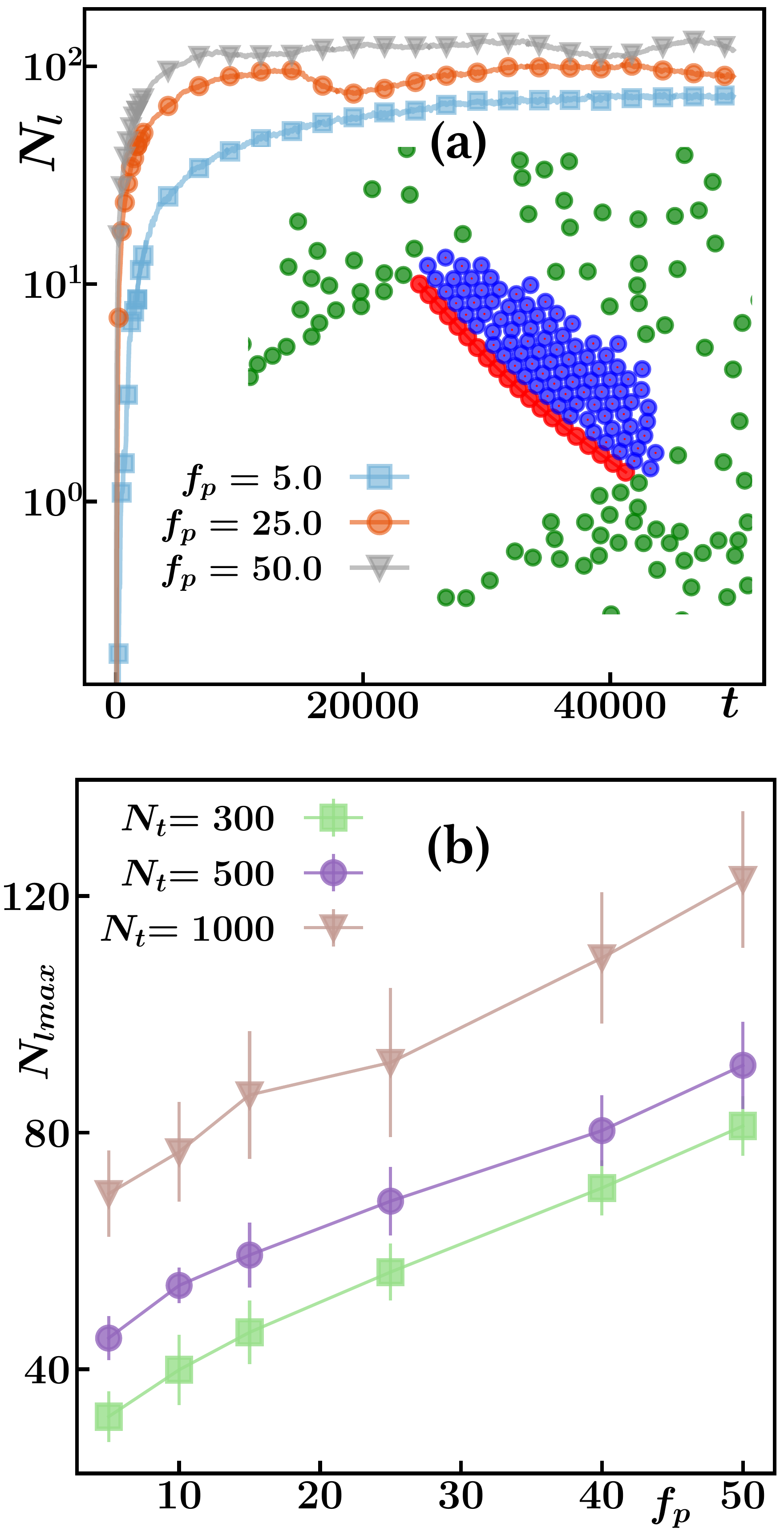}
	\caption{ (a) The number of load particle ($N_l$) as a function of time for different values of $f_p$, $N=20$. Initially the filaments collects less particle and then gradually   the rate of collection increases and eventually the it saturates  to $N_{lmax}$.  Inset: Snapshot of a portion of the simulation box. Identified the loaded particles using the cluster algorithm is shown in blue. The rest of the passive brownian spheres colored green. ( Here, $N=20, N_t=1000, f_p=25.0$). (b) The dependence of  $N_{lmax}$ on the value of $f_p$ and the number of background brownian spheres $N_t$. is  pair correlation function for different values of $f_p$ for the rod length $N=20$ and $N_t=1000$.  }
	\label{fig:figsnapshot-Nlt-gr}
\end{figure} 
 
To understand the particle collection when $N_t >>N$, we randomly keep the Brownian particles inside the simulation box along with the single propelling polymer.  Since  $N_t$ is large, we employ a cluster-finding algorithm to identify the number of particles ($N_l$) that are swept by the polymer at a given time $t$.   To begin with, we find a set of all the particles  ($G_n$) within a cut-off distance $r_c<2^{(1/6)} \sigma$ from any other particles. This set contains any pair of particles within the above cut-off, not just close to the polymer. The next step is to identify the particles in close proximity to the polymer from the elements of  $G_n$ by applying the same cut-off distance from each monomer of the polymer. We group all such particles within the cut-off from any of the monomers of the polymer in the first neighbour list. Since each of these particles sit in the void between the two nearest monomer beads of the polymer,  they form a layer next to the polymer. Nearest neighbours for all the particles in the layer are searched from $G_n$ again. By repeating the process of finding new neighbours and their nearest neighbours, we build a cluster containing $N_l$ particles. We present such a cluster segregated from the $G_n$,  where the blue particles are the identified load ~($N_l$) particles (only a portion of the simulation box is shown) and green particles are free passive particles, in the inset of  Fig~\ref{fig:figsnapshot-Nlt-gr}(a). Due to the absence of any attractive interaction between the rod and Brownian particles during the motion and also due to the collision of the other free particles, we observe that a small portion of loaded particles gets free near the end of the polymer. 

  In Fig.~\ref{fig:figsnapshot-Nlt-gr}(a), we present the number of particles collected by the rod as a function of time. During the initial stage, the number of particles trapped by the rod is very small. Then $N_l$ increases rapidly as the rod finds more and more particles during its translational motion. Then finally, it saturates to a maximum number $N_{lmax}$. From Fig.~\ref{fig:figsnapshot-Nlt-gr}(a), it is clear that the number of collected particles depends on the propulsion force. With low propulsion, the rod encounters fewer particles, and the number of particles collected is less.  $N_{lmax}$ is estimated from the saturated region of $N_l-t$ plot.  Due to the dependence of $f_p$ on $N_{lmax}$, in order to decipher more information, we varied the number of surrounding beads $N_t$ while keeping the polymer length constant. We consider three different cases of $N_t {~(300, 500, 1000) }$ to study the effect of passive beads. When there are more passive particles in the surrounding medium, the probability of encounter between the rod and the $N_t$ increases and more particles are collected, see Fig.~\ref{fig:figsnapshot-Nlt-gr}(b). This suggests that the rod can collect more particles in a denser environment, even with the lower propulsion strength. In all the three cases $N_{lmax}$ linearly increases with $f_p$, see  Fig.~\ref{fig:figsnapshot-Nlt-gr}(b). However, at much higher $f_p$, the filament slightly deforms from the straight conformation, and it generates a minute curvature (as seen in the inset of Fig.~\ref{fig:figsnapshot-Nlt-gr}(a)), thus reducing the number of particles collected (in comparison to the expected value from the low $f_p$ regime). Thus we have two different slopes for small and large $f_p$ values, even though the linear relation holds in both regimes. 

 \begin{figure}
	\includegraphics[width=0.45\textwidth]{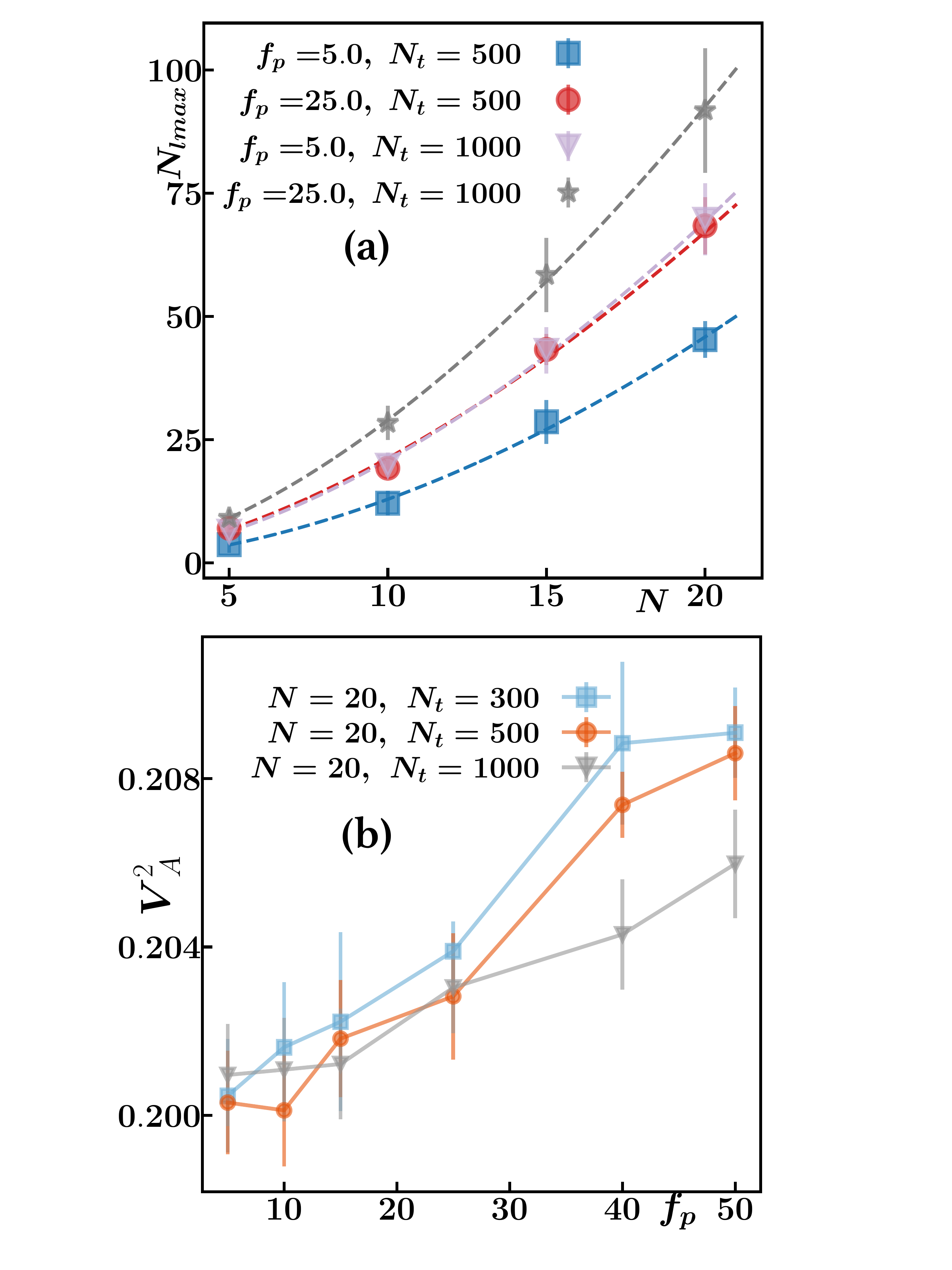}
	\caption{ (a) Snapshot of a portion of the simulation box. Identified load using the cluster algorithm is shown in blue. The rest of the passive brownian spheres colored green. Here $N=20, N_t=1000, f_p=25.0$  (b) The number of load particle ($N_l$) as a function of time. Initially the filaments collects less particle and then gradually   the rate of collection increases and eventually the it saturates  to $N_{lmax}$. Inset shows the depends of  $N_{lmax}$ on the value of $f_p$ and the number of background brownian spheres $N_t$. is  pair correlation function for different values of $f_p$ for the rod length $N=20$ and $N_t=1000$.  }
	\label{fig:figNlmax_veloctiy}
\end{figure}

Compared to spherical beads or active filaments propelling along its long axis, transverse filaments have more 'active area' for collecting cargo particles. When the length of the filament is increased,  it directly increases the 'active area' of collection. Now we examine how increasing the polymer length relates to the maximum number of particles collected. We chose two different  $f_p$ and $N_t$ values to examine the effect of polymer length as shown in Fig.~\ref{fig:figNlmax_veloctiy}.

 Experiments on the sweeper rod suggested that the trapped cargo particles arrange like a triangular lattice-like structure~\cite{janussweeper-21}. Since the monomer and passive particles are of the same size,  we assume that cargo particles sit at the void between monomers when it touches the rod. From a geometrical argument, for a stiff polymer having $N$ monomers,  one can argue that to make a complete triangular lattice-like structure in $2D$ on top of it, the maximum number of passive beads needed is ($N_{lmax}$)= $N(N-1)/2$, $N_{lmax}\sim \frac{N^2}{2}$. In Fig~\ref{fig:figNlmax_veloctiy}a, we are showing the dependence of  $N_{lmax}$ on polymer length for different values of $f_p$ and $N_t$.   For all the values of $f_p$ and $N_t$ the  $N_{lmax}$ scales as $~\mathcal{A}N^{s}$, (dashed fits in ~\ref{fig:figNlmax_veloctiy}a). Here $\mathcal{A}$ is a prefactor, and the value is found to be less than $0.5$ and  the value of the exponent, $s$, lies between $1.5<s<1.9$. The reduced values of the exponent and prefactor from the expected values can be ascribed to the filament slight deformation  when the particles are collected. The curvature of the filament also helps to reduce the loss of collected particles from the edge of the filament. The overlapping curve suggests that a slower filament in a dense environment and a faster filament in a dilute background can collect the same number of particles. Thus  maximum loaded particles depends clearly on the number of passive beads, length of polymer and its propulsion strength.  
 
 Now we investigate, what happens to the velocity of the rod when it collects the maximum number of particles ($N_{lmax}$). It is expected from the theoretical prediction that when there is an infinite load,  the velocity of the rod reduces to  zero~\ref{fig:vcm-no-load}. This is because the rod will experience an increased drag force due to the additional particles it is carrying, which will cause it to slow down. However,  in our simulation, we see that the polymer continues its motion even after collecting a large number of particles (See SI movie). The motion of the polymer could be due to the constant propulsion force applied to the rod, and the velocity does not necessarily reach zero but rather is reduces to a value close to the contribution from thermal velocity, the other contributing factor to the velocity of the rod.  See the figure ~\ref{fig:figNlmax_veloctiy}b. The tendency of saturating to a terminal value of velocity is visible in figure \ref{fig:vcm-no-load}b.   
  
\begin{figure}
	\includegraphics[width=0.45\textwidth]{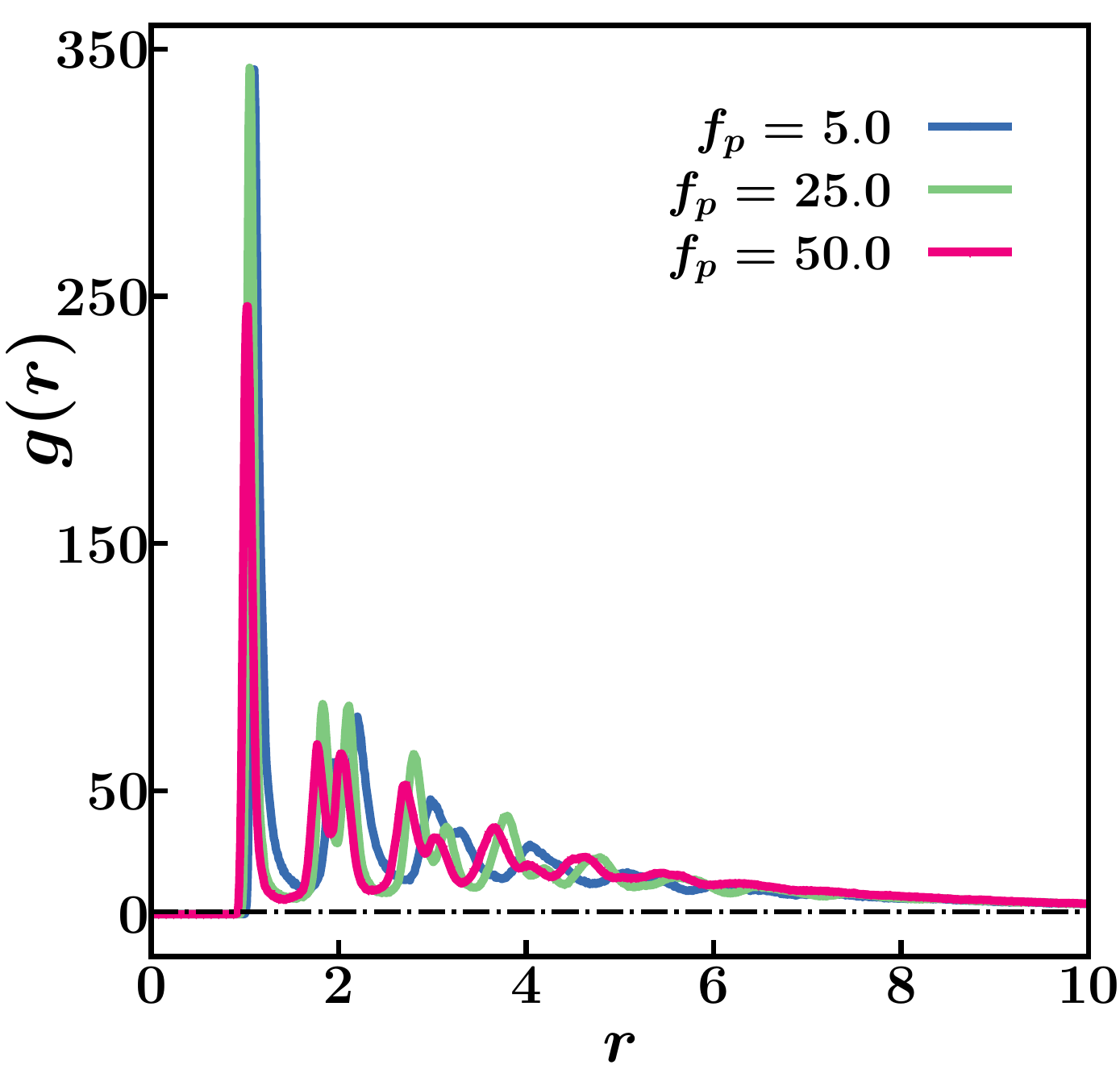}
	\caption{ g(r) for different values of $f_p$. We used $N=20$ and  $N_t=1000$. The location of the peaks  corresponds to  the lattice points of a closed packed structure formed by the trapped particles.}
	\label{fig:gr}
\end{figure}

Lastly, we examine how the passive cargoes are arranged when the sweeper traps them. Experiments demonstrated that passive particles are arranged like a triangular particle raft; however, detailed information related to the structure is unavailable.  In our simulation, when the particles get collected, the repulsive interaction between the rods and passive particles dominates, and cargo arrange like a  "triangular-lattice" like structure. To get a qualitative picture of the structure of the maximum load, we examine the pair correlation function of the collected particles. We measure $g(r)= \frac{A}{2\pi  r\Delta r }\langle\frac{dn(r)}{N_l}\rangle$,  and is presented in Fig.~\ref{fig:gr}. $dn(r)$  is averaged over the distribution of neighbours around each of the $N_l$
particles (between the distance $r\pm\Delta r/2$), and then over the configurations of the system for a large number of consecutive times ($\langle\cdot\cdot\cdot\cdot\rangle$). We consider the $A$ as $L_x L_y$, the area of the simulation box. While the first peak corresponds to the average nearest neighbour distance $\sigma$, the second and third represent the position of the average next-nearest neighbour and third-nearest neighbour located at $1.73\sigma$ and $2\sigma$ respectively. As $f_p$ is increased, the polymer deformation creates domains or defects in the arragment of loaded particles, and these peaks shift to slightly lower values. It suggests that arrangement of the cargos is influenced by the strength of sweeping motion  of the active polymer and the interactions between the cargos themselves. The smooth decay of $g(r)$  to zero reflects that the measurement is only based on the maximum loaded particles.

\section{IV. Summary and outlook}
In this paper, we studied a mixture of  passive particles  and a transversely propelling polymer using the Langevin dynamics simulation. Our work is inspired by the recent experiment on sideways moving bimetallic rod transporting poly-styrene beads as cargo~\cite{janussweeper-21}. In our simulation, an active  polymer experiences local propulsion forces in the direction perpendicular to its contour, leading to a sideways propulsion of the polymer. Cargos are modelled as passive Brownian particles undergoing thermal fluctuations. We assume steric repulsion between the passive particles and the polymer chain. The steric interaction between the particles prevents any crossing between them. We considered a dry limit, where we ignored any effect of fluid flow. 

We have shown that when the propulsion is perpendicular to the long axis of the body,  it can be employed as a "sweeper" to collect passive particles. Sideways propelling polymer moves by a constant active force, and the velocity of the polymer is reduced as it "sweeps" the passive beads. The velocity reduction is associated with the drag of the particles. However, when the polymer collects the maximum load, we show that it attains a terminal velocity, and its value is close to the contribution from thermal fluctuations. The number of particles collected by the road increases as a function of time, reaching a maximum value after a transient time. The maximum number of trapped particles scales with the   length of the sweeper as it increases the "area" of collection. Additionally, we demonstrate that the propulsion strength and the number of Brownian particles are other determining factors for the maximum load. This suggests that moving slower in dense environments and moving faster among fewer particles have a similar effect on particle collection. Similar to experiments, 
we show that the aggregates of the collected particles arrange in a triangular closed, packed state.  Moreover, we show  that cargos collected by the rod  arrange in a triangular, closed, packed state, similar to what has been observed in experiments.  The structural arrangement of the trapped particles depends mainly  on the propulsion  force and the interaction between the passive particles. 

The model considered here is a simplified version, as we neglected the effect of hydrodynamic flows and electrophoretic attraction between the rods and beads as in the experiments. 
However, we emphasise that our model captures several generic features of the experiments and quantitatively mimics the experimental outcome. Introducing more complexity to the model imparts further investigation and might impact the maximum load and the scaling.   We hope that our study will motivate further experimental and theoretical research on non-specific cargo trapping using sideways moving filament  when conventional methods fails. 

In the present work, we considered only one type of passive particle for simplicity. In real systems, the unwanted species (or cargos) may be  of different sizes and shapes or may have various interactions with each other and with the sweeper. Thus, it will be interesting to investigate the possibility of considering a  passive mixture with different sizes, shapes and interactions for representing cargo in the future. This could provide much insight into  information such as the maximum load the polymer can transport, the arrangement of the collected particles and whether they can get segregated during the trapping process etc.

Finally, we emphasise that a static chevron-shaped wall with a medium apex angle is preferred for the complete trapping of active objects~\cite{Lowen-trapping-PRL-12,kaiser2013capturing}. Also, it was shown that a boundary with a sharp local curvature acts as the nucleation site for collective trapping. While for the transverse sweeper collecting passive particles, a wider 'angle' for the sweeper is favourable for accumulating a large number of particles. To ascertain the influence of the angle (the corresponding local curvature), which may play a role in the efficiency of particle trapping and transport, we can employ transverse filament, with faster moving ends inducing curvature, for particle transport~\cite{prathyusha_softmatter-22}. 

The next step is incorporating the model with controlled transport and cargo delivery capability. Different such approaches could involve controlling the propulsion direction, tuning the surface pattern of the substrate, or manipulating the interaction between the sweeper and cargo particles. Future studies could focus on incorporating different propulsion mechanisms (Eg: chemical) for the transverse propelling polymer and study how it can be used for specific or non-specific cargo transport. We hope our work will motivate exciting avenues for these future extensions to better understand how to effectively and efficiently transport and deliver cargo particles in different types of systems and environments.





\section{V. Acknowledgements}
KRP would like to acknowledge K. Ramesh Kumar for many useful discussions. 

%

\end{document}